\documentclass[11pt,a4paper]{article}
\usepackage{jcappub}
\usepackage{amssymb}
\usepackage{graphicx}
\usepackage{amsmath}
\usepackage{amsfonts}
\usepackage{hyperref}
\usepackage{subfig}

\title{Soft gamma-ray repeaters and anomalous X-ray pulsars as highly magnetized white dwarfs}

\author[a]{Banibrata Mukhopadhyay\footnote{Corresponding Author.}}
\author[b]{and A. R. Rao }

\affiliation[a]{Department of Physics, Indian Institute of Science, 
Bangalore 560012, India}

\affiliation[b]{Tata Institute of Fundamental Research,
Mumbai 400005, India}

\emailAdd{bm@physics.iisc.ernet.in}
\emailAdd{arrao@tifr.res.in}

\abstract{
We explore the possibility that soft gamma-ray repeaters (SGRs) and anomalous X-ray pulsars (AXPs)
are powered by highly magnetized white dwarfs (B-WDs).
We take a sample of SGRs and AXPs and provide the possible parameter space in mass, radius, and
surface magnetic field based on their observed properties (period and its derivative) and the
assumption that these sources obey the mass-radius relation derived for the B-WDs.
The radius
and magnetic field of B-WDs are adequate to explain energies in SGRs/AXPs as
the rotationally powered energy. In addition, B-WDs also adequately explain the perplexing radio
transient GCRT~J1745-3009 as a white dwarf pulsar.
Note that the radius and magnetic fields of B-WDs are
neither extreme (unlike of highly magnetized neutron stars) nor ordinary (unlike of magnetized white dwarfs,
yet following the Chandrasekhar's mass-radius relation (C-WDs)). In order to explain SGRs/AXPs, while
the highly magnetized neutron stars require an extra,
observationally not well established yet, source of energy, the C-WDs
predict large ultra-violet luminosity which is observationally constrained from a strict upper limit.
Finally, we provide a set of basic differences between the magnetar and B-WD hypotheses for
SGRs/AXPs.
}

\arxivnumber{ }

\begin{document}
\maketitle

\section{Introduction}

Soft gamma-ray repeaters (SGRs) and anomalous X-ray pulsars (AXPs) are astronomical objects 
which exhibit pulsations and their properties are different from the rotation powered
radio pulsars and accretion powered X-ray pulsars. SGRs/AXPs are, as of now,
most popularly hypothesized to be isolated, spinning down, highly magnetized neutron stars (NSs)
(magnetar model) \cite{thomdun92}. According to
this model, emission is powered by the energy stored in strong magnetic fields 
of NSs with the surface field $B_s \sim 10^{14}-10^{15}$G. 
Such strong 
magnetic fields, however,  have not been detected from observed data yet and alternate models like fast rotating white dwarfs (WDs) have been proposed \cite{pac,usov,usov2}.

AXPs are distinguished from X-ray binaries
by their narrow period distribution, soft X-ray spectrum, faint optical counterparts, and
long term spin-down. SGRs, on the other hand, are observed with their bright and short bursts and hence
they are considered as a subclass of gamma-ray bursts. Based on their persistent 
X-ray counterparts, SGRs were found to be very similar to AXPs and hence they
are often classified together, as is in the present work. The rotational 
period of the AXPs/SGRs, as of now, lies in a narrow range ($2-12$ s) compared
to that for the ordinary pulsars. Moreover, their generally large spin-down rates,
strong outburst energies and giant flares make them different from ordinary pulsars.

Typical rotationally powered energy in AXPs/SGRs is $\sim 10^{32}-10^{34}$ ergs s$^{-1}$ while their
X-ray luminosity $\gtrsim 10^{34}$ ergs s$^{-1}$, which
rules them out to be considered as rotationally powered NSs. 
A NS of radius $10$km with $B_s\sim 10^{14}$ G and central field $B_c\sim 10^{16}$ G
can have magnetic energy $\sim 10^{48}$ ergs s$^{-1}$, which could produce the luminosity 
$\sim 10^{36}$ ergs s$^{-1}$, 
in its typical age (if 
they are associated with supernova remnants or young clusters of massive stars).
Such a high field, based on the propagation time of magnetic instability in the NS 
surface with Alfv\'en speed, explains the short duration of initial spike in giant flares.
Furthermore, the strong field explains the confinement of the hot plasma required
for the subsequent tail with a softer spectrum pulsating at the NS rotation period.
Other phenomena being explained due to the existence of large magnetic field include
short bursts in almost all AXPs/SGRs with peak luminosity exceeding by a few orders of 
magnitude the Eddington limit for a NS and high frequency quasi-periodic oscillations (QPOs).

While the above description generally makes the foundation of magnetar model 
concrete, there are certain shortcomings in it. First of all, as of now, there 
is no evidence for the strongly magnetized NS --- as strong as required for the magnetar
model. 
The inferred/measured strongest possible field, in some occasions, has
been only $\sim 10^{12}$ G \cite{meregh13,rea,jap}.  
Second, recent Fermi observations
are inconsistent with the predicted high energy gamma-ray emissions in the magnetars
\cite{tong}. Third, inferred upper limit of $B_s$ for some magnetars, e.g. SGR~0418+5729,
is quite smaller than the field required to explain observed X-ray luminosity. 
Fourth, it has been shown (e.g. \citep{allen,ferrario,woosley}) that
the attempt to relate magnetars to the energies of the supernova remnants or
the formation of black holes is not viable.
Among other inconsistencies, there are
radio pulsars discovered with inferred $B_s$ overlapping with those 
of AXPs/SGRs, but without any signature of magnetically powered emissions like bursts/flares.
Nevertheless, by arguing all of them to be rotating high-field NSs with respective magnetic axes having 
different orientations with respect to their rotation axes, they were attempted to be unified \cite{bingmag}.
In the presence of strong magnetic fields, the high-field pulsars can have active inner 
accelerators while the AXPs cannot, revealing different observed emission characteristics
among themselves. 
However, recently discovered NSs with relatively low inferred dipole field 
exhibit outburst properties similar to those in AXPs/SGRs.
These observations imply that a high magnetic dipole moment is not a mandatory condition
for a magnetar. 

In order to remove these shortcomings, 
at least a part, 
AXPs/SGRs have been again re-argued to be magnetized WDs \cite{mane}, following the 
idea originally proposed more than two decades back \cite{pac,usov} (see also \cite{ost}). Due to their larger radius ($\sim 10^4$ km
for a typical WD), the rotationally powered energy for WDs
could be quite larger than that for NSs. Hence, these authors attempted 
to explain the energy released in AXPs/SGRs through the occurrence of glitch and 
subsequent loss of the rotational energy. Indeed, the possibility of glitch in WDs, as starquakes, with mass 
around the Chandrasekhar limit \cite{chandra} was shown earlier \cite{usov}
which explained the mean spin-down rate observed for 1E~2259+586.  
However, this WD based model (hereinafter C-WD) is challenged by the observed short
spin periods (e.g. \cite{meregh13}). In addition, due to larger radius, they should exhibit larger 
ultra-violet (UV) luminosities,
which, however, suffer from a deep upper limits on the optical counterparts (e.g. \cite{durant,meregh13}) of
some AXPs/SGRs, e.g. SGR~0418+5729 (see, however, \cite{ruf}).

Recently, Mukhopadhyay and his collaborators, in a series of papers, have proposed
for the existence of highly (as well as very highly) magnetized WDs \cite{mpla,prd,ijmpd1,prl,apjl,ijmpd2}
with mass significantly super-Chandrasekhar. 
$B_s$ of such WDs (hereinafter B-WDs)
could be as high as $10^{12}$ G and $B_c$ could be $2-3$ orders of magnitude higher (see \cite{nord}).

Any new idea, when proposed, generally is tested with a simplistic model first. Once, the results
based on a simplistic model show promise to explain observations and/or experiments, then more realistic
self-consistent models, introducing more sophisticated physics, are introduced in order to
fine-tune the original model. Without being an exception, Mukhopadhyay and collaborators 
have also followed the same tactics
to develop their super-Chandrasekhar WD model (see \cite{mukh}, which discusses the evolution 
of this topic so far).

These authors have, so far, approached towards this mission through the following steps. First, they have
considered most simplistic, spherically symmetric, very highly magnetized WDs in the Newtonian
framework, assuming the magnetic field to be constant or almost constant throughout (or modeling,
as if, the inner region of WDs) \cite{prd}. 
To assure stability of such WDs, the authors assumed that the large magnetic field in them
is {\it tangled/fluctuating} in a length scale larger than the quantum length scale such that
the average field and hence corresponding magnetic pressure is much smaller than the matter
pressure modified by the Landau effects \cite{prl,revisit}.

However, it has been speculated in those work itself that with very high fields
the self-consistent consideration of deformation of WDs would reveal a similar
super-Chandrasekhar mass at lower fields. In the same model framework, they have also shown that
magnetized WDs altogether have a new mass-limit, $80\%$ larger than the Chandrasekhar-limit \cite{prl},
in the same spirit as the Chandrasekhar-limit was obtained \cite{chandra}.

Afterwards, the authors have removed both the assumptions: the Newtonian description and spherical symmetry
(e.g. \cite{jcap14,jcap15a}).
Note that magnetized WDs could be significantly smaller in size compared to their
conventional counter-parts \cite{prd,prl} and, hence, general relativistic effects therein may not be negligible.
Thus, based on a full scale general relativistic magnetohydrodynamic (GRMHD) description \cite{pili},
they have explored more self-consistent WDs which are ellipsoid and have revealed
similar stable masses, as obtained in the simpler framework, but at smaller fields \cite{jcap15a,sathya}, as
speculated earlier \cite{prd,revisit}.
In fact, in the later work \cite{jcap15a,sathya}, the authors are able to show that depending on the 
field profiles (which were chosen self-consistent in accordance with other conditions/equations) and 
hence magnetic pressure gradient and magnetic density, maximum mass 
of a B-WD could be even slightly higher than that proposed earlier \cite{prl}, but at high fields 
(not {\it very high} fields).

In order to understand, how to acquire the strong magnetic fields in B-WDs,
the present authors argued \cite{apjl} for a possible evolutionary scenario by which super-Chandrasekhar WDs 
with high magnetic fields could be formed by accretion on to a commonly observed magnetized
WD, invoking the phenomenon of flux freezing. Based on a varying accretion rate scenario,
they showed that a highly super-Chandrasekhar B-WD could be formed in the time scale
of $\sim 2\times 10^7$ years from a $0.2M_\odot$ WD with surface field $\sim 10^9$ G. 
The idea is, as WD accretes matter, its magnetic field amplifies as a consequence of
the increase in (central) density, hence gravitational force,
due to the contraction in size (via flux freezing theorem) of WD \cite{cumm}.
Nevertheless, other ways (including possible dynamo based mechanisms) of generating such strong 
fields, at least partially, cannot be ruled out
(however note that depending in field profiles, a very strong field is not necessarily
needed to form a super-Chandrasekhar B-WDs). The authors also
showed that the estimated number of super-Chandrasekhar B-WDs, governed from cataclysmic variables 
(CVs), is consistent with the 
observed rate of peculiar type~Ia supernovae, which could be their ultimate fate 
(at least one of the possible fates).

Such WDs are already shown to help 
in explaining peculiar, over-luminous type~Ia supernovae whose progenitor masses
are necessarily super-Chandrasekhar. Nevertheless, these B-WDs are smaller or significantly 
smaller in size, depending on the field geometry, compared to their ordinary counterparts (e.g. polar with $B_s\sim 10^9$ G).
Typically their radius can just be an order or two magnitude(s) higher than that of a NS.
As the surface temperatures of WDs with the change of magnetic fields 
are not expected to be significantly higher \cite{ferrar} (in fact could be lower \cite{mukul}), 
smaller the radius smaller the luminosity 
of the WD is. Therefore, B-WDs should be consistent
with the UV-luminosity ($L_{UV}$) cut-off in AXPs/SGRs. Moreover, their typical $B_s$ is consistent 
with observation,
but adequate to explain AXP/SGR energies as rotationally/spin-down powered energy, unlike magnetars.

Furthermore, B-WDs could be  adequate candidates to explain certain 
peculiar radio pulsars/transients as white dwarf pulsars (WDPs), 
e.g. GCRT~J1745-3009 \cite{zg03}, which is otherwise thought to be
the prototype of a hitherto unknown class of transient radio sources.

Hence, in the present paper, we explore AXPs/SGRs as B-WDs. Note that the
idea to explore B-WDs as AXPs/SGRs has already been presented 
by independent groups \cite{emil,rao}. Although the evolution of B-WDs could be 
by accretion, they may 
appear as AXPs/SGRs at the exhaustion of mass supply after significant evolution. 
Such WDs' $B_s$ and $R$ combination can easily explain AXPs/SGRs as 
rotationally powered WDs. 
All the machineries implemented in the magnetar model 
can be applicable for B-WDs as well, however, with a smaller $B_s$ which
is physically more viable. 
As the ranges of radii and magnetic fields of B-WDs lie in between those of highly magnetized 
NSs and C-WDs --- neither very extreme nor ordinary, we propose these ranges to be explored
extensively in understanding related observations (not restricted to AXPs/SGRs and WDPs only), 
in particular the ones which remain unresolved yet.
Therefore, the present work aims at initiating this venture.

The paper is organized as follows. In the next section, we outline the model to explain magnetized WDs 
as rotating dipoles. Subsequently, in \S\S 3 and 4, we explore a few sources of SGR/AXP and GCRT~J1745-3009
as B-WDs respectively. In \S 5, we present the basic differences between the magnetar and B-WD models.
Finally, we end in \S 6 with summary and implications.

\section{Modelling magnetized white dwarfs as rotating dipoles}

Following standard electrodynamics \cite{jack}, the rate of energy loss
from a rotating, magnetized compact star, assuming it to be a rotating dipole, is \cite{arc}
\begin{equation}
\dot{E}_{\rm rot}=-\frac{4\Omega^4\sin^2\alpha}{5c^3}|m|^2,
\label{edot}
\end{equation}
when the variation of dipole moment $m$ arises due to the inclination of magnetic axis 
with respect to the rotational axis of the star by the 
angle $\alpha$, $\Omega$ is assumed to be the angular frequency
of the star at the surface, $c$ the light speed, $\mu_0$ the vacuum permeability. Now 
the dipole nature of the magnetic field
can be expressed as
\begin{equation}
B=\frac{2 |m|}{ R^3},
\end{equation}
when $R$ is the radius (or average radius if it is spheroid) of the star. However, the above energy loss rate can be defined as 
the rate of change of rotational kinetic energy of the star with moment of inertia $I$ so that
\begin{equation}
I\Omega\dot{\Omega}=\dot{E}_{\rm rot},
\end{equation}
which leads to 
\begin{equation}
B_s=\sqrt{\frac{5c^3IP\dot{P}}{4\pi^2R^6\sin^2\alpha}}~G,
\label{bs}
\end{equation}
when $P$ is the rotational period and $\dot{P}$ the period derivative.
This is the upper limit of $B_s$. 
As $P$ and $\dot{P}$ for AXPs/SGRs (and WDPs) are known from observation, $B_s$ can be computed from
a given mass-radius ($M-R$) relation for rotating B-WDs when $\alpha$ is a parameter. Note that 
only that $M$ and $R$ (or equatorial radius $R_e$) are the approprite set from the $M-R$ relation whose surface 
angular velocity ($\Omega_{eq}$) corresponds to the observed $P$. Once $B_s$ is estimated for
an observation, the rotational/dipole energy $E_{\rm rot}$ stored in the star can be computed. This
furthermore quantifies the maximum energy stored in it, if there is no other source as adopted
in the magnetar model.

We explore the possibility to explain the origin of high energy phenomena in AXPs/SGRs and WDPs 
by $\dot{E}_{\rm rot}$ and $B_s$ of B-WDs --- there is
no need to invoke extraordinary, yet observationally unconfirmed, sources of energy. This is possible 
because B-WDs have larger
$I$ (due to larger $R$) than NSs, revealing larger $m$, which is however small enough to produce 
UV-luminosity. 

The $M-R$ relations of B-WDs depend on the magnetic field \cite{prd,prl,jcap15a,sathya}. 
It has been self-consistently found by GRMHD simulations \cite{jcap15a,sathya} that field 
decreases from the central region to the surface at least in $2-3$ orders of magnitude.
As $B_c$  
plays the major role in holding the mass \cite{prd}, an $M-R$ relation corresponds
to the strict value(s) of $B_c$ too. However, $B_s$ is weakly constrained. A range of $B_s$ 
corresponds to very similar $M$ and $R$ for a given $B_c$, as long as $B_c/B_s\gtrsim 10^3$. 
Nevertheless, from the solution of stellar structure, a given $M$ and $R$ corresponds to a given $B_c$ and 
$B_s$ \cite{prd,prl}, as well as given $\Omega_{eq}$ \cite{sathya} for rotating stars. 
On the other hand, for a given observation ($P$ and $\dot{P}$), a particular set of  
$M$, $R$ and $\Omega_{eq}$ corresponds 
to a particular $B_s$ from equation (\ref{bs}), which has to be same/similar to the value of $B_s$ corresponding 
to the magnetic field profile giving rise to chosen $M$ and $\Omega_{eq}$ in the first place. 
This helps in removing the apparent degeneracy in $B_s$
with the solution of stellar structure. For a given $M-R$ relation, only the combinations of $M$, $R$ and 
$\Omega_{eq}$, which satisfy the above constraint of $B_s$, are useful for explaining AXPs/SGRs and WDPs. 
This outlines the rule followed here. 

\section{Explaining AXPs/SGRs}

We consider nine AXPs/SGRs to estimate the range of possible parameters of B-WDs. We consider
two cases separately. First, we consider models with 
highly magnetized (with $B_c\lesssim 5\times 10^{14}$ G) B-WDs considering anisotropic effects 
of magnetic field self-consistently formulated in GRMHD simulations \cite{jcap15a,sathya}. 
Subsequently, we assume the B-WDs to be very highly magnetized (with $B_c\gtrsim 10^{15}$ G)
and follow the respective approximate models \cite{prd,prl}. 

\subsection{B-WDs with high magnetic fields}

\begin{figure}[h]
\vskip 0.3in
   \centering
\includegraphics[angle=0,scale=0.4]{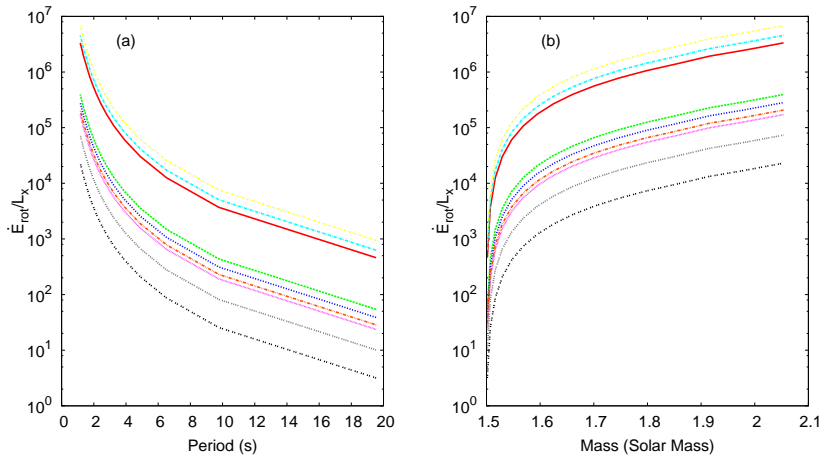}
\vskip0.5cm
\caption{The ratio of rate of rotational energy released to X-ray luminosity as a function of 
(a) spin period, and (b) mass, for B-WDs
when from the top to bottom various curves correspond to 1E 1547-54, 1E 1048-59, 
SGR 1806-20, SGR 1900+14, SGR 0526-66, SGR 1822-1606, 1E 1841-045, SGR 0418+5729 and 1E 2259+586.
For other details, see Table~1.
}
\label{hmedot}
\end{figure}

We consider a typical $M-R_e$ relation, as shown in Fig. 20 of \cite{sathya}, described
for poloidal magnetic field profiles with $B_c=3.1\times 10^{14}$ G, $R_e$ ranging in
$1534-1586$ km and corresponding $R_p/R_e$ in $0.82-0.55$ ($R_p$ being the polar radius). 
If the surface temperature is assumed to be $\sim 10^4$ K, then $L_{UV}\sim 10^{29}$ ergs/s.  
However, a more self-consistent computation reveals $L_{UV}$ to be much smaller in the 
presence of field considered here \cite{mukul}.
The observed values of $P$, $\dot{P}$ and $L_x$ for nine SGRs/AXPs are given in Table~1.
Figure \ref{hmedot} shows that $\dot{E}_{\rm rot}$ computed based on our model, with a fixed $\alpha=15^\circ$, 
in the range $1\le P/{\rm sec}\le 20$ is several orders of magnitude larger than observed $L_x$ 
for each source. Note that only one point in each curve (corresponding to fixed $\dot{P}$) 
in Fig. \ref{hmedot}a corresponds to the respective source with observed $P$, when different
$M$ in the $M-R_e$ relation corresponds to different $\Omega_{eq}$ and, hence, different $P$. 
From Fig. \ref{hmedot}b we can retrieve $M$ and, hence $R_e$, corresponding to the 
respective source. It can be mentioned that the $\sin\alpha$ factor in equation (\ref{edot})
(and hence other equations) does not have much importance, because a wind component can also 
spin down a pulsar. In fact, the detailed $\alpha$-dependence can be different in different models
(see, e.g., \cite{alpha1,alpha2}). Hence, we do not intend to constrain $\alpha$-dependence in our model
and the computations are done for a particular, fixed value of $\alpha$, unless stated otherwise.

\begin{figure}[h]
\vskip 0.3in
   \centering
\includegraphics[angle=0,scale=0.4]{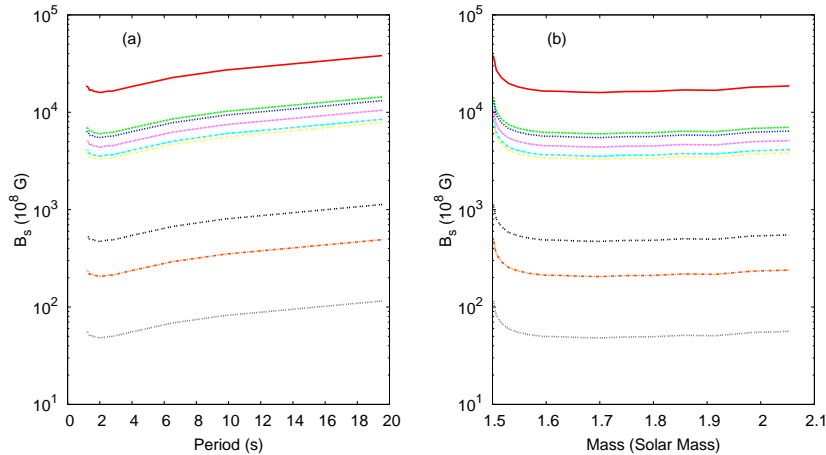}
\vskip0.5cm
\caption{Surface magnetic field as a function of (a) spin period, and (b) mass, for B-WDs
when from the top to bottom various curves correspond to SGR 1806-20, SGR 1900+14, 
SGR 0526-66, 1E 1841-045, 1E 1048-59, 1E 1547-54, 1E 2259+586, SGR 1822-1606 and SGR 0418+5729.
For other details, see Table~1.
}
\label{hmbs}
\end{figure}

Figure \ref{hmbs} shows $B_s$ for the respective sources. As above, only one point in each curve 
corresponds to the respective source with known $P$. The values of $B_s$, along with their
$B_c$, are confirming these B-WDs to be excellent storage of magnetic energy. The computed
$B_s$ values turn out to be much higher compared to those in C-WDs (see Fig. \ref{rad}b, 
which will be discussed furthermore in \S 3.2). Hence, the $M-R_e$ relation adequately
explains all nine AXPs/SGRs without requiring an extra-ordinary source of
magnetic energy. 

\subsection{B-WDs with very high magnetic fields}

Some of the 
properties of chosen AXPs/SGRs, particularly useful for the present modelling, are listed in Table~1. 
Like in \S3.1, we primarily consider
$\alpha=15^\circ$ which is not observationally much constrained either. However, in some
sources, $\alpha$ must be smaller (if no other source of spin down is considered) in order to explain observed 
data better according to our theory.
Indeed, $\alpha$ for SGR~0418+5729 has been argued to be very small \cite{tong-alfa}. For
such sources, a range of $\alpha$ is tabulated, when smaller the $\alpha$, larger the allowed mass range of 
the WD is. However, a larger $\alpha$ reveals a smaller $L_{UV}$ which
argues the WDs to be more difficult to observe. We also assume the surface temperature 
of WDs to be $T_{UV}\sim 10^4$ K and set their radius of gyration assuming the WDs to be
semi-solid sphere/ellipsoid.

\begin{table}[h]
\vskip0.2cm
{\caption{ Various observational and theoretical parameters of AXPs/SGRs: $P$, $\dot{P}$,  $L_x$ are 
observed values and inputs and $\alpha$, minimum of $L_{UV}$ are outputs of our very highly magnetized 
B-WD model, discussed in \S 3.2. See, http://www.physics.mcgill.ca/$\sim$pulsar/magnetar/main.html}}
{\centerline{}}
\begin{center}
\begin{tabular}{|c|c|c|c|c|c|c|}

\hline
AXPs/SGRs & $P$ &  $\dot{P}$ & $L_x$ & $\alpha$
& $L_{UV}$$_{\rm min}$ & $L_{UV}$$_{\rm min}$ \\
 & (s) & $(10^{-11})$ & $(10^{35}$ ergs s$^{-1}$) & (degree) & (ergs s$^{-1}$)& (ergs s$^{-1}$)\\
 & &  &  &  &B-WD & C-WD\\
\hline

1E~1547-54 & $2.07$ & $2.32$ & $0.031$ & $5-15$ & $5.7\times10^{28}$&$4.8\times10^{29}$\\
\hline
1E~1048-59 & $6.45$ & $2.7$ & $0.054$ & $5-15$ & $3.5\times10^{26}$& $9.2\times10^{29}$\\
\hline
1E~1841-045 & $11.78$ & $4.15$ & $2.2$ & $15$ &$1.6\times10^{28}$&$1.7\times10^{30}$\\
\hline
1E~2259+586 & $6.98$ & $0.048$ & $0.19$ & $2-3$& $3.4\times10^{26}$ &$1.5\times10^{29}$\\
\hline
SGR~1806-20 & $7.56$ & $54.9$ & $1.5$ & $15$ &$3.4\times10^{26}$&$3.5\times10^{30}$\\
\hline
SGR~1900+14 & $5.17$ & $7.78$ & $1.8$ & $15$ &$8.6\times10^{28}$ & $1.3\times10^{30}$\\
\hline
SGR~0526-66 & $8.05$ & $6.5$ & $2.1$ & $15$ &$6.4\times10^{27}$ & $1.7\times10^{30}$\\
\hline
SGR~0418+5729 & $9.08$ & $5\times 10^{-4}$ & $6.2\times 10^{-4}$ & $1-5$ &$3\times10^{28}$ &$1.8\times10^{29}$\\
\hline
SGR~1822-1606 & $8.44$ & $9.1\times 10^{-3}$ & $4\times 10^{-3}$ & $1-5$ &$3.4\times10^{26}$ &$8\times10^{28}$\\
\hline

\end{tabular}

\end{center}
\end{table}

\begin{figure}[h]
\vskip 0.3in
   \centering
\includegraphics[angle=270,scale=0.4]{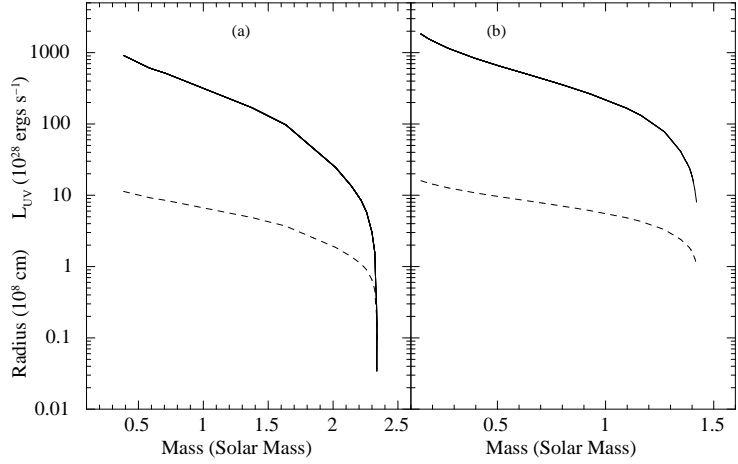}
\vskip0.5cm
\caption{UV-luminosity (solid line) and radius (dashed line) as functions of 
mass for (a) B-WDs, and (b) C-WDs.
}
\label{rad}
\end{figure}

Figure \ref{rad}a shows the $M-R$ combinations for B-WDs, which exhibit
$100\lesssim\dot{E}_{\rm rot}/L_x\lesssim 10^7$ from equation (\ref{edot}) for the  
AXPs/SGRs listed in Table~1, explaining them well as rotational powered pulsars. The corresponding $L_{UV}$ appears to
be as small as $\sim10^{26}$ ergs s$^{-1}$ (see Table~1). Figure \ref{rad}b along with Table~1, however, shows that 
the $M-R$ combinations for the C-WDs exhibit $1-4$ orders of magnitude 
higher $L_{UV}$.

\begin{figure}[h]
\vskip 0.3in
   \centering
\includegraphics[angle=270,scale=0.4]{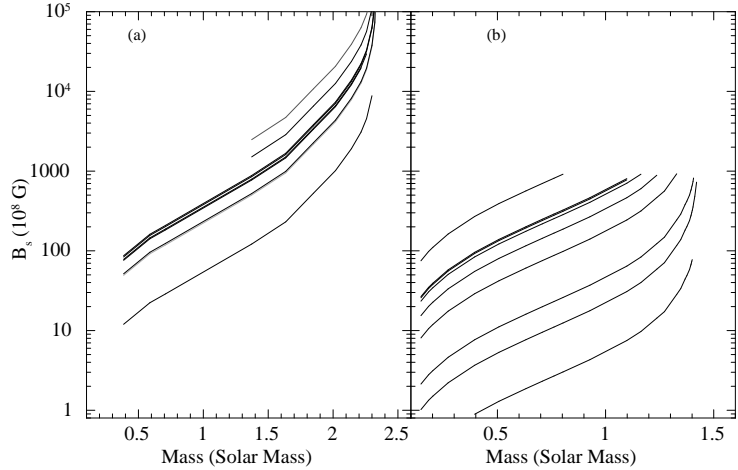}
\vskip0.5cm
\caption{Surface magnetic field as a function of mass for (a) B-WDs when
from the top to bottom various curves correspond to SGR~1806-20, 1E~1048-59, 
SGR~0526-66, 1E~1841-045, 1E~1547-54, SGR~1900+14, 1E~2259+586, SGR~1822-1606, SGR~0418+5729, (b) C-WDs
when from the top to bottom various curves correspond to SGR~1806-20, SGR~0526-66,
1E~1841-045, SGR~1900+14, 1E~1048-59, 1E~1547-54, 1E~2259+586, SGR~1822-1606, SGR~0418+5729.
}
\label{bb}
\end{figure}

\begin{figure}[h]
\vskip 0.3in
   \centering
\includegraphics[angle=270,scale=0.3]{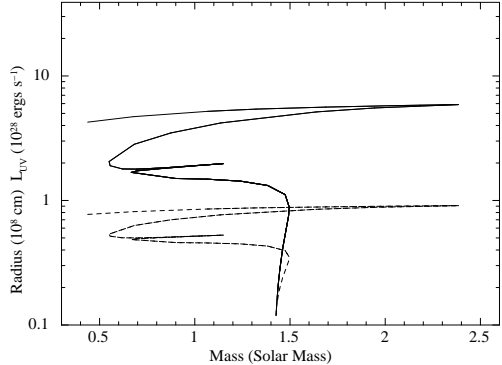}
\vskip0.5cm
\caption{UV-luminosity (solid line) and radius (dashed line) as functions of mass for a particular
central magnetic field $7\times 10^{15}$ G. 
}
\label{radcb}
\end{figure}

Figures \ref{bb}a and \ref{bb}b show that $B_s$ of the B-WDs is quite stronger
compared to that of the C-WDs. 
This reveals
that B-WDs are better storage of rotational/spin-down/magnetic energy. They
naturally explain AXPs/SGRs without requiring an extra-ordinary source of magnetic energy. 
Generally higher the $B_s$ and $B_c$, higher the $M$ is, which corresponds to a lower $R$
and hence a lower $L_{UV}$. However, as shown earlier \cite{prd}, depending on the central
density, a B-WD of lower $R$ with lower $M$ ($\sim 1.5 M_\odot$) can be formed. Therefore,
for a given $M$, $L_{UV}$ for B-WDs could be smaller than that for C-WDs. Figure \ref{radcb} shows that 
a B-WD of $M\sim 1.45 M_\odot$ can have $L_{UV}\sim 10^{27}$ ergs s$^{-1}$ which is significantly 
smaller than the smallest possible $L_{UV}$ for a C-WD. 

Generally (see Table~1), the expected  $L_{UV}$-s of B-WDs are two 
orders of magnitude lower than that of C-WDs, except a few cases where they are just a few factors lower. 
It should be interesting to concentrate on these sources and measure their UV flux and spectrum, 
simultaneously
with the X-ray spectrum, so that contributions from  putative companions could be subtracted and 
the real blackbody emission from the compact objects could be measured and radius be estimated.

\section{Explaining GCRT~J1745-3009}

GCRT~J1745-3009 is a transient radio source in the direction of the galactic center, which exhibited five peculiar
consecutive outbursts at $0.33$ GHz with a period of $77.13$ minutes. Zhang and Gil \cite{zg03} argued it to be
a WDP with a period of $77.13$ minutes within $0.8$ kpc. Later on, by color-magnitude analysis, 
Kaplan et al. \cite{kaplan} showed that WDs with typical temperatures ($5000-20,000$ K) and radii 
($5000$ km) could not be this much close to us and would be at least at around $2$ kpc. This
challenged the idea of the source to be a WDP. Now in the framework of very slowly rotating B-WDs 
(see Figs. 17 and 18 of Ref. \cite{sathya}) with $B_s\sim 3.3\times 10^{11} - 2\times 10^{12}$ G,
corresponding $R\sim 1580 - 500$ km (depending on whether the magnetic field is high or very high) 
and central density $\sim 10^{10}$ gm cm$^{-3}$, we revisit all
the calculations, e.g. radius of polar cap and unipolar potential drop therein,
done by previous authors \cite{zg03}, and find them to be consistent with WDP idea when the unipolar potential drop     
is at least one order of magnitude smaller than that in radio pulsars. However, the emission altitude
turns out to be in accordance with radio pulsars for $\alpha\sim 30$ and $\sim 15$ for high and very high
fields respectively. Furthermore, the sum of mean free paths for 
electrons to produce inverse-Compton gamma-ray photons and for the photons to attenuate turns out
to be larger than the radius of B-WDs, which rules out the possible pair productions and explains
why GCRT~J1745-3009 is dormant before and after the bursting cycles. The maximum gamma-ray/X-ray
flux appears to be only a factor of 4 larger than that obtained by Zhang and Gil \cite{zg03} for the 
same parameters for very highly magnetized B-WDs, which is still quite small to detect. While for highly
magnetized cases it appears to be $\sim 100$ times larger than that obtained by previous authors,
it is very consistent with the X-ray flux upper limit of $\sim 5\times 10^{-10}$ ergs s$^{-1}$ cm$^{-2}$ 
\cite{hym}. Moreover, the solid angle of radio emission is quite 
unknown and is a free parameter in order to reduce gamma-ray/X-ray flux furthermore.
Interestingly, $\dot{P}\sim 10^{-14}$ for such B-WDs
which is difficult to measure. 

Finally, from the condition of radio luminosity not exceeding the spin-down luminosity reveals
the distance of the source to be $\lesssim 8.5$ kpc (for the same solid
angle chosen by the previous authors), for high magnetic fields, which is much larger than that predicted by Zhang and Gil \cite{zg03} and
in accordance with the lower limit predicted by Kaplan et al. \cite{kaplan}. For very high magnetic fields, 
the distance turns out to be smaller, but still at least $\sim 1.6$ kpc which is larger than that
predicted by Zhang and Gil \cite{zg03}. As shown in \S 3,  
such B-WDs are significantly cooler and hence being further away than that predicted by 
Zhang and Gil \cite{zg03}, its optical flux will be dimmer to evade detection. This strongly supports
the source to be a WDP.

\section{Magnetar versus B-WDs}

  The magnetar hypothesis is around for quite some time and several observational
features of SGRs/AXPs are worked out under the premise of this hypothesis. 
While attempting to explain new observations, sometimes new factors in the magnetar
hypothesis are invoked, which, although not proven, are quite plausible. For example,
 the period derivatives in magnetars are often found to vary, sometimes related to
other parameters, like, luminosity and temperature, and these are explained
by invoking the torque exerted by the magnetosphere (see, e.g., \cite{kaspi2}). 
Many of these explanations can possibly  be tailored to suit the
B-WD hypothesis, presented in this work. 

Instead of embarking on such an adventure, we 
 enumerate here the most fundamental and basic differences between the magnetar
model prevailing in the literature and the B-WD hypothesis presented in this paper, to
explain the properties of SGRs/ AXPs. Future observations can be tuned to 
pin down these differences which will help in either identifying the exciting new
objects called B-WD or it will strengthen the magnetar hypothesis.

\subsection{Narrow period range and rarity of the objects}  
     
    The narrow period range and the rarity of the SGRs/AXPs is explained in the
magnetar hypothesis as due to the peculiar conditions required to produce the
extremely high magnetic field. They form presumably with a very high spin,
and, due to the extremely high magnetic field, spin down very fast to the
observed range of periods. At longer periods, they are too weak to be detected.
The B-WDs, on the other hand, are rare objects and their formation may require still 
rarer constraint of being able to accrete matter from a low mass companion.
The lower period range would be too fast for a WD, whereas at longer periods
they could be undetectable. 

   The observational consequence of these would be as follows. In the magnetar hypothesis, there
is a rare possibility of detecting SGRs/AXPs with fast rotation (significantly 
less than a second), whereas such objects would not be possible in the B-WD hypothesis.
Furthermore, in the magnetar hypothesis, there should be a large number of high magnetic
field long period NSs with quite weak emission. In the B-WD hypothesis, however,
slowly rotating high magnetic WDs could be detected as the progenitors of peculiar
Type Ia Supernovae.

\subsection{Origin and age}

  Magnetars necessarily have to be young and should be associated with a young
supernova remnant (SNR). If one measures the space velocities of SGRs/AXPs, from the measured age, 
one should be able to identify the recent SNR for each and every SGR/AXP. The B-WDs
would be preferentially seen in dense stellar regions (so that the peculiar high magnetic
field WDs capture a low mass companion and increase their core magnetic field by
accretion), but they need  not be associated with SNR.

\subsection{Companion and relic accretion disks}

 In the B-WD hypothesis, one of the methods to generate higher $B_c$ is
by accretion and the resultant flux freezing in WDs. Hence if B-WDs are 
the central sources in SGRs/AXPs, some of these objects may retain
a low mass companion and/or relic accretion disks.  
 Being a recently born NS in a supernova, no such companion or relic
accretion disk should be found in SGRs/AXPs, in the magnetar hypothesis.

\subsection{Explaining giant flares}

One of the motivations to introduce the magnetar concept was
to explain the huge amount of energy released in repeated bursts,
in particular giant flares, in SGRs.
In the B-WD model, the SGR/AXP systems are perhaps old systems, mildly powered  
by relic accretion and mostly powered by rotation. Hence, a continuous decay
of magnetic field as well as increase in mass (due to accretion) is possible
(and, hence, decrease in radius due to stronger gravity and increase in
magnetic field due to flux freezing) in them. This is expected to result in frequent re-adjustment
of the high magnetic field and its gradient within the star, as well as 
equation of state (for very high magnetic field cases). This could furthermore 
result in outbursts in SGR/AXP systems in the B-WD hypothesis.

\subsection{Size}

One of the most fundamental differences in the two models is the size of the
compact object. In the magnetar hypothesis, they are NSs with a size
of $8 - 10$ km. In the B-WD hypothesis, they can have a wide range of size ranging
from a few tens of km to a few thousands of km. 
A detailed modelling and understanding
of the X-ray spectral components should be able to resolve the source size.
Indeed, detailed X-ray spectroscopy of AXPs has revealed
the presence of thermal components in the X-ray spectrum (see,
e.g., \cite{enoto}) and it is shown that the inferred sizes of the
emission region are sometimes lower than a km, demonstrating small
hot spots in the emission, rather than emission from the full
surface of the compact object. The detection of UV/optical emission of sources,
on the other hand, points towards the emission from the full surface
of the compact object. Hulleman et al. \cite{hull}
already pointed out that the optical data of AXP~4U 0142+61
are consistent with a hot WD.

Hence, the multi-wavelength data of AXPs need to be re-looked
in the perspective of optical/UV emission coming from the
surface of a B-WD and X-rays coming from hot spots, to derive sensible
source parameters.

\subsection{Non-detection of gamma-rays and radio emissions}

As per the B-WD model, the spin energy of the WD is used to power
the X-ray emission with a mechanism similar to that seen in
rotation powered pulsars (RPPs), with, however, different size and magnetic
fields. Hence, many of our understanding of RPPs must be applicable to 
AXPs/SGRs, as per the B-WD model. RPPs are normally seen in radio wavelengths 
and among the thousands of RPPs discovered in radio wavelengths, only a 
few dozen are seen in X-rays and fewer still in gamma-rays. It is found that 
$L_x$ scales as $B^2/P^4$ (when $B$ is average magnetic fields) and the 
gamma-ray luminosity, $L_\gamma$, scales as $B/P^2$ (e.g. \cite{hard}).
Hence, only the young RPPs with lower periods and higher magnetic fields 
are seen in X-rays and gamma-rays. 

Now B-WDs could have larger values of $B$ than ordinary NS
RPPs, which also implies their narrower polar caps. The source XTE J1810-197,
however, was seen in radio wavelengths \cite{camilo}
and the authors pointed out that employing
RPP typed mechanism cannot be excluded for the handful of
known magnetars, because of their long periods implying
small active polar caps and narrow beams which may miss the observers
for random orientations. Hence the rare detection of radio pulsations
in AXPs/SGRs, as compared to NS RPPs, is consistent with the high magnetic 
field assumed for B-WDs.

The non-detection of AXPs/SGRs in gamma-ray energies 
is already in conflict with the outer gap model in the magnetar
scenario \cite{tongsong}. Since gamma-ray
luminosity scales as $B$, the lower-than magnetar field for B-WDs explains
the lower $L_\gamma$ for AXPs/SGRs in the B-WD model.

To summarize, in the B-WD scenario, the magnetic field could be
higher than NS RPPs to have narrow beams and field could be
lower than magnetar model to have lower $L_\gamma$. A deeper search,
however, for gamma-ray emission in AXPs/SGRs and a detailed modelling using 
the B-WD model parameters will certainly help in distinguishing/refining
the B-WD model.

\subsection{Capability of measuring magnetic fields by latest experiments}

Recently launched ASTROSAT and Hitomi satellites are capable of detecting 
hard X-rays of energy upto $\sim 100$ keV. Hence, they are expected to 
observe more sources with the capabilities of wide band spectroscopy. Now the electron
Cyclotron resonance energy of absorption spectrum is to be $E_c=11.6~(B_s/10^{12}~G)$ keV.
Therefore, a compact object with $B_s\sim 10^{12}-10^{13}$ G is possible to be observed
by ASTROSAT/LAXPC and could be identified as a B-WD. However, for a NS based magnetar
with $B_s\gtrsim 10^{14}$ G, $E_c\gtrsim 1000$ keV, which is quite beyond the scope
of these satellites.

\section{Summary and Implications} 
We have demonstrated important applications of recently formulated B-WDs \cite{prd,ijmpd1,prl,ijmpd2}.
The present work indicates a possibility of
wide application of B-WDs in modern astrophysics. The idea that AXPs/SGRs need sources
of energy other than  accretion is certainly inevitable, but the hypothesis
that they are highly magnetic NSs, although attractive, did not neatly fit in 
with furthermore observations (unlike other ideas in Astrophysics, like, spinning NSs
as radio pulsars and accreting compact objects as X-ray binaries, which quickly  
established themselves as paradigms). 
In other words, while there is a more standard model as of now, namely magnetar, to explain AXPs/SGRs,
alternate models must be explored keeping shortcomings in magnetar model in mind and 
here we have explored one such model developed recently, namely B-WDs.

Apart from the limitations of the magnetar model discussed
earlier, the model does not have natural explanation of several observed features, like,
the narrow period range (however, within the framework of the magnetic
field decay models, it was attempted it explain \cite{narP}) and several orders of magnitude 
range in period derivative, lack of
large proper motion, lack of supernova identification for all AXPs/SGRs etc. Hence, it
is very important that other possible explanations for the AXP/SGR phenomena, at least a part 
of them, need to be
seriously explored. The B-WD concept is an extremely attractive alternate for AXPs/SGRs. 
This is because of its range of mass and radius to satisfy different  observations.

Recently, the source 1E~2259+586 has been reported to exhibit an anti-glitch \cite{kaspi}.
While the presence of glitch has been explained in the theory of NS, an anti-glitch is difficult
to argue under the same framework. We, however, can speculatively describe it in B-WDs as follows. 
As we argued earlier, an equilibrium B-WD will have a larger interior field which decays (starting from the core-crust
boundary) down away from the center 
\cite{apjl}. However, B-WDs are expected to encounter
continuous mass-loss, along with the decay of its angular velocity and magnetic field exhibiting rotational power. 
Now, when the surface field decays below a critical value, the stiffness of the field profile increases
significantly enough to produce an outward force arising from the additional gradient of magnetic pressure,
say $\sim 2\times 10^{11}$ ergs cm$^{-3}$ at a radius $\sim 600$ km with a density $\sim 10^9$ gm  cm$^{-3}$. This in turn leads to the 
increase of the stellar radius in equilibrium and, from the
conservation of angular momentum, a decrease of the frequency with an anti-glitch $\sim -4.5\times 10^{-8}$ Hz as observed. 
Furthermore, above a critical time scale, the density will decrease appreciably to decrease the inner magnetic field
due to continuous mass-loss, when the outer field has been already smaller. This in turn decreases the stiffness 
of the field profile and hence the gradient of magnetic pressure. As a result, this leads to the gravitational
power to be suddenly stronger to exhibit glitch. This cycle may continue depending on the competitive powers
between mass-loss and rotational energy extraction rates. The glitch, however, can also be occurred due to the
loss of angular velocity alone and subsequent star-quake, in the same model platform, as is argued in the 
standard NS picture.

Another important application of B-WDs has been explaining the peculiar radio transient GCRT~J1745-3009 
as a WDP. While earlier authors indeed argued it to be a WDP, subsequently it was ruled out   based on a more accurate prediction of its distance in the framework of a C-WD. However,
considering it to be a B-WD furthermore opens up its possibility of a WDP, due B-WDs' larger magnetic field
and smaller radius which could predict it to be low luminous and further away and hence dimmer to evade detection.

Now, in order to verify the usefulness of B-WDs, it is important to refine X-ray spectroscopic tools to
measure the mass and radius of the compact objects in AXPs/SGRs. 
In the calculations described in this
work, we could only derive a range of possible magnetic field related to the corresponding mass and radius 
for a given source based on the observed $P$ and $\dot{P}$.
If, observationally, some parameters could be tied to either
mass or radius (even magnetic field, e.g. by electron Cyclotron absorption line, which would be in the 
X-ray regime, unlike the gamma-ray regime of NS based model), then unique solutions and hence definite predictions can be made based on the B-WD model.
In the larger astrophysical context, it is important to examine various observations in the light of the
existence of B-WDs and their creation by the phenomena of flux-freezing. For example, the fact that 
in the SDSS survey, the magnetic WDs are found to have, on an average, larger mass compared to the
non-magnetic WDs \cite{ferrar} can be explained naturally if we assume that the magnetic WDs are
generated by accretion (and the resultant contraction in radius and flux freezing). Similarly, magnetic
WDs in CVs have higher field strength below their period gap (which are
evolved systems), again, comes naturally from the fact that evolved WDs, due to accretion, should have
larger mass, lower radius, and higher magnetic field. Finally, a precise mass measurement of evolved
CVs should identify a few of them as B-WDs. \\

\acknowledgments
The authors would like to thank Bing Zhang for comments and suggestions.

\end{document}